\documentclass{article}

\usepackage{PRIMEarxiv}

\usepackage[utf8]{inputenc} 
\usepackage[T1]{fontenc}    
\usepackage{hyperref}       
\usepackage{url}            
\usepackage{booktabs}       
\usepackage{amsfonts}       
\usepackage{nicefrac}       
\usepackage{microtype}      
\usepackage{lipsum}
\usepackage{fancyhdr}       
\usepackage{graphicx}       
\graphicspath{{media/}}     

\usepackage[numbers]{natbib}

\usepackage{multirow}
\usepackage{float}
\usepackage{subfig} 
\usepackage{amsmath}
\usepackage{mathtools}
\usepackage{bm}
\usepackage{multirow}
\usepackage{enumitem}
\usepackage[ruled,vlined,linesnumbered,noend]{algorithm2e}
\usepackage[table,xcdraw]{xcolor}

\usepackage{combelow}

\title{Advanced Medical Image Representation for Efficient Processing and Transfer in Multisite Clouds}

\author{
  Elena-Simona Apostol$^1$ and Ciprian-Octavian Truic{\u{a}}$^1$\\
  $^1$University Politehnica of Bucharest, Bucharest, Romania \\
  \texttt{ciprian.truica@upb.ro, elena.apostol@upb.ro}
}

\begin{document}
\maketitle

\begin{abstract}
An important topic in medical research is the process of improving the images obtained from medical devices.
As a consequence, there is also a need to improve medical image resolution and analysis.
Another issue in this field is the large amount of stored medical data~\citep{Mazzamuto2018}.
Human brain databases at medical institutes, for example, can accumulate tens of Terabytes of data per year.
In this paper, we propose a novel medical image format representation based on multiple data structures that improve the information maintained in the medical images. The new representation keeps additional metadata information, such as the image class or tags for the objects found in the image. We defined our own ontology to help us classify the objects found in medical images using a multilayer neural network.
As we generally deal with large data sets, we used the MapReduce paradigm in the Cloud environment to speed up the image processing.
To optimize the transfer between Cloud nodes and to reduce the preprocessing time, we also propose a data compression method based on deduplication. We test our solution for image representation and efficient data transfer in a multisite cloud environment. Our proposed solution optimizes the data transfer with a time improvement of 27\% on average.
\end{abstract}


\maketitle

\section{Introduction}\label{sec:introduction}

Medical image analysis is a crucial component in the process of establishing a diagnosis, planning treatments, or evaluating surgical procedures.
An important issue in the context of handling large medical data is that today there are no standardizations for the efficient processing of images from different types of medical devices.
In this context, the image representation must be universal and must provide a very good resolution.
As an example, consider multiple anatomical images from a patient produced with Magnetic Resonance Imaging (MRI) technology and representing different stages in tumor evolution.
Image analysis techniques are required to assess, for example, the malignancy of the tumor or plan the required treatment, using these images.
These medical images, in general, have different dimensions and intensities, making this a very challenging task.
Also, in order to have a better representation, additional information obtained from the image acquisition device must be inserted into the image.
For example, nowadays, a Positron Emission Tomography-Computed Tomography (PETCT)  image can have more than 4GB.
A patient that has cancer, with a high chance of relapse, can have a history of $1$ PETCT/year and up to $3$ NMRs/year for a five years period.
The diagnosis process involves lots of parameters, but the medical history is a key aspect.
A medicinal doctor would have to store more than $50GB$ of data just for a patient.
Another key aspect is the availability of the data: if the patient needs to see multiple medicinal doctors for a differential, the data must be available for all of them.
For an analysis of the evolution of different types of cancer, the researchers must deal with Terabytes of data that need to be classified and analyzed.
Therefore there is a need for a new image format that enables the comparison of images from different devices and improves the scalability of the diagnosis process.

The research questions we are trying to answer are:
\begin{itemize}
    \item[$Q_1$.] Can we provide a standard in image representation to improve the medical images (obtained from different devices) processing and transfer in a distributed environment?
    \item[$Q_2$.] Can a new metadata-enhanced image representation standard help in efficiently retrieving sufficient information on a patient’s disease and assist the medicinal doctor in putting a diagnosis?
\end{itemize}

In order to allow medicinal doctors to gain information from the different registration types, we propose a new image format representation that is invariant to affine transformations and that stores additional data.
The metadata inserted in the image is related to the registration criteria: images can be 2D, 3D, mono modal, multimodal~\citep{Wang2013}, for one patient, or used in an atlas. 
The use of statistical atlases in medical applications has increased during the past years.
They can help physicians to discover patterns for different anatomical structures and to understand how these structures change over time or react to medications.
In order to represent a universal format, we focused on the main aspects that define an image and are of relevance for a medicinal doctor, i.e., the pixels (location, the color that can lead to contrast, hue, luminosity, etc.) and the content (objects in the image, defining a region of interest, finding that region in a different image, etc.).
As a result, the size of the proposed image is considerably higher than that of the original image.
Thus, we also propose an efficient solution for processing and storing these images using the MapReduce paradigm and Cloud Storage.
The MapReduce paradigm offers the possibility to distribute and scale large amounts of data.

The new proposed format is stored using a Quadtree~\citep{Kirichek2018QT} based representation that allows the processing and manipulation of images in a shorter time than other types of image representations while also requiring less storage space.
Considering the distributed Cloud environment, one of the main drawbacks is that different parts of the image need to be transferred to different nodes.
The transfer will increase the computational time (considering the limitations: bandwidth, throughput, etc.) and the price. 
To address this, we propose a new transfer method based on data deduplication. 
Data deduplication is defined as a data compression technique used for eliminating data copies. 
An image consists of many irrelevant data points that will only slow down the image processing algorithm. 
We address this problem by defining a Merkle tree structure on top of the Quadtree layers.
Our experiments show the efficiency of the proposed solution in a multisite cloud distributed environment.

To sum up, the main contributions of this paper are as follows:
\begin{itemize}
    \item[$C_1$.] Propose an image representation that can preserve geometrical transformations and image resolution while applying the different global transformations.
    \item[$C_2$.] Propose a hybrid data representation and a deduplication-based transfer algorithm that allow us to speed up the data transfer and reduce the cost of the system.
    \item[$C_3$.] Propose an efficient solution for processing and storing medical images using the MapReduce paradigm and Cloud Storage.
    \item[$C_4$.] Enhance the images' metadata with information from preprocessing images using clustering algorithms and neural networks.
    \item[$C_5$.] Define our own ontology to help us classify the objects from the medical images and to determine correlations between the objects of interest.
\end{itemize}

This article is structured as follows:
Section~\ref{sec:limitations} presents the main limitations of the technologies used in medical imaging.
Section~\ref{sec:related_work} presents state-of-the-art solutions for image deduplication and representation.
Section~\ref{sec:solution} discusses the methodology for our novel deduplication system and presents the new image representation invariant to affine transformations.
Section~\ref{sec:implementation} presents the methodology and implementation details for our solution.
Section~\ref{sec:experiments} shows the experimental validation of the proposed system.
Finally, Section~\ref{sec:conclusions} presents our conclusions and hint at future work.

\section{Limitations of Medical Imaging}\label{sec:limitations}
In the medical imaging area, where there is a wide range of equipment that generates images in various formats, defining an image representation standard to improve medical image processing and data transfer is critical.
A viable solution has to consider, in addition to numerous performance measures, the current limitations in the field.
In this section, we will discuss the challenges and limitations of the medical imaging field.

The ability to improve the resolution limits of sensors using super-resolution algorithms~\citep{Robinson2011} has shown significant progress in the area of photographic imaging.
By far in this area, the majority of applications using super-resolution technology were for either consumer or defense-type applications.
Recently, researchers integrated super-resolution algorithms in different imaging medical applications.
Trying to improve the resolution of medical images, the researchers took into consideration the differences between photographic imaging and medical imaging.

Usually, when a medical image is taken, a controlled source of illumination is often used during the image acquisition.
The strong illumination provokes a higher Signal-to-Noise Ratio (SNR)~\citep{Priyadharshini2013}.
In many cases, the strong illumination can cause damage to the exposed tissue, similar to what happens when we are exposed to the sun for a long period.

The imaging speed is another factor that has to be taken into consideration.
In order to avoid patients' movement that would minimize the associated imaging artifacts, the image must be taken faster than a photo-shooting.
Furthermore, the image processing artifacts are much less tolerable in medical images than in photographic applications.
In a fortunate manner, medical imaging systems operate under highly controlled environments with highly similar objects.
The images' quality can be improved by leveraging prior knowledge about the anatomy or biology to improve image quality~\citep{Merino2007}.

Finally, the majority of medical imaging applications involve creating images from radiation propagation through three-dimensional objects.
Thus, while the final images are two-dimensional, they represent some form of projection through a three-dimensional volume.

The following paragraphs are concerned with presenting different medical imaging techniques and their main limitations.

\paragraph{\textbf{Low Radiation Digital X-ray Mammography}} 
Today's digital detectors cannot shrink pixel sizes to increase resolution without sacrificing the SNR measurement~\citep{Karasev2013}.
To maximize image resolution, the researchers have explored the idea of digitally combining multiple low-dosage images, each containing spatial shifts.
Providing high-resolution imagery required sophisticated, nonlinear reconstruction techniques to address the extremely low SNR of the captured images, but using an improved super-resolution algorithm solved the problem~\citep{Robinson2011}.

\paragraph{\textbf{Optical Coherence Tomography}}
Optical Coherence Tomography (OCT) systems provide noninvasive yet high-resolution in-vivo images of retinal structures.
The images can be used to define imaging biomarkers of the onset and progression of many ophthalmic diseases~\citep{Roychowdhury2013}.
When used in real-time, the OCT system skips frames.
While the patient has his/her eyes scanned, he/she can voluntary or involuntary blink which changes sight directions.
Skipping frames will render the OCT system ineffective.
If a new image format would be used that would preserve more information or enhance the parallel processing of the digital input signal, the ability of the system to generate new data will improve.

\paragraph{\textbf{PET and SPECT}}
Positron emission tomography (PET) uses coincidence detection to image functional processes.
PET images can be viewed in comparison to computed tomography scans to determine an anatomic correlate.
Single Photon Emission Computed Tomography (SPECT) is a 3D tomographic technique that uses gamma camera data from many projections and can be reconstructed in different planes.
The main problem with both positron emission tomography and SPECT is they are limited by the amount of radiation that the patient is exposed to.
During a year, a patient can have only a limited number of computed topographies, that are calculated based on different factors: patient's age, body structure, weight, height, medical condition, etc.
Using the existing protocol, Digital Imaging and Communications in Medicine (DICOM), it is difficult to compare two images that were taken during a longer period of time for a patient from different devices, or even two images that were taken with the same device, but calibrated differently.

\section{Related Work}\label{sec:related_work}
In this section, we analyze the current solutions for image deduplication and representation.

\subsection{Image Deduplication}
As the nature of medical data is huge in size, Big Data management environments are required for storing and preprocessing such large volumes of information~\citep{Ahmad2021}.
Cloud based multi-site distributed applications are required to facilitate high data availability, data redundancy, and partition tolerance in case of failure.
These applications are employing image deduplication techniques to decrease storage cost~\citep{Sujatha2020,SureshAnand2020} and increase security~\citep{Marwan2017} as images are encrypted and decrypted using hash values based on image content~\citep{Gang2015,Pintilie2016}.
To ensure source authentication and to protect the existence of medical image watermarks, new encryption techniques that use deduplication are required~\citep{Priya2019}.
Furthermore, currently, there is a high requirement for secure, cost-efficient architectures which allow both patients and medical practitioners to improve the disease management and treatment process, while making the data access and sharing process among the practitioners who treat the same patients seamlessly.
\cite{Arka2014} propose a framework that increases accessibility to medial data by employing secure image compression and decompression techniques that incorporate in-memory caching and pre-fetching to access both the records and their metadata.

Convergent encryption is used to securely eliminate duplicate copies on encrypted data which supports only exact data deduplication.
To address this strict requirement, \cite{Li2015} propose a privacy-preserving fuzzy image deduplication scheme SPSD (secure perceptual similarity deduplication scheme).
By measuring image similarity over encrypted data, the SPSD eliminates duplicated images by employing a perceptual hash algorithm to produce image signatures for similarity comparison.
This approach is robust to perceptual-persisting modifications, e.g., compression, flipping, and resizing.
Using the cryptographic hash-mapping function, SPSD manages to decrease storage and bandwidth costs while providing secure and efficient image deduplication and secure sharing among patients and the practitioners that treat them.

The Locality-sensitive hashing (LSH) algorithm is used to create a privacy-preserving near-duplicate image data detection scheme in order to store both the ciphertexts of image data and image-feature metadata vector in cloud systems~\citep{Wu2018}.
During retrieval of the near-duplicate image data, the LSH algorithm is used to generate the image-feature query token for the metadata vector.
The query token is then used to return the encrypted result by employing the privacy-preserving near-duplicate image data detection scheme.
Besides preserving privacy, the LSH-based schema also achieves query correctness, i.e., the content returned is correct and lightweight, i.e., the near-duplicate detection is done with the minimum communication and computation costs.

As medical images contain highly sensitive data, new proper encryption algorithms that facilitate data aggregation and compression are required.
\cite{Wang2019} propose a novel Compressive Sensing (CS) scheme.
CS uses homomorphic aggregation to offer data confidentially, image sampling to facilitate digital processing and fast retrieval, and compression to decrease storage costs.
\cite{Yadav2020} use a Message Locked Encryption and Convergent Encryption calculation for the deduplication strategy to decrease storage and increase computational execution costs.
The proposed framework scrambles and re-encodes client data simultaneously.
Furthermore, they use a square level deduplication strategy to further minimize storage costs.

SafePHR~\citep{Fu2021} is a secure and efficient medical data service with application aware deduplication in fog-to-multicloud encrypted storage.
The system combines low-latency and high-safety to ensure fog-to-multicloud cooperative multicloud encrypted storage.
It also uses disaster recovery mechanisms to enhance eHealth data management.
SafePHR employs a deduplication schema to improve data traffic and space efficiency, while for encryption it uses the PHR variants of convergent encryption.
Thus, SafePHR ensures data confidentiality while minimizing the storage space overhead using deduplication.

\subsection{Image Representation}
Currently, there are different models for efficiently representing medical images, e.g., based on spatial relationship, vector based, or Quadtree.

Several solutions propose a representation based on spatial relationship model.
This type of model is generally used for more efficient content-based image retrieval.
In this model, the spatial relationships among different objects are used to further handle multiple objects in an image.
In Table~\ref{tab:0}, we present the main advantages and disadvantages of Spatial Relationship Image representation.
To design such a model, \citet{Ahn2021} proposes a new spatial guided self-supervised clustering network (SGSCN) for medical image segmentation. 
The model introduces multiple loss functions to aid in grouping image pixels that are spatially connected and have similar feature representations. 
It iteratively learns feature representations and clustering assignments of each pixel in an end-to-end fashion from a single image. 
\citet{Ren2022} propose a medical image super-resolution method based on semantic perception transfer learning, for benefiting clinicians in disease diagnosis.
Their approach also takes into account spatial relationships between objects and builds a semantic feature extraction network as well as an image description generation network.
 It also uses image and text modal data to learn transferrable, high-level semantic characteristics.

Some solutions for medical image analysis (e.g., ~\cite{Sadeghpour2023, Cao2021vector}) assume a Vector based representation.
The properties of being editable and scalable make this a viable solution when working with large amounts of data.
Also, it is easy to render data and save it to another format file or vector format, with good results.
One of the main disadvantages of using this model is that image reconstruction may take considerably longer compared with a bitmap file of equivalent complexity because each image element must be drawn individually and in sequence.
Also, it is restrictive as it cannot store extremely complex images (e.g., if color info is paramount, it varies on a pixel-by-pixel basis).
The two most important vectorization techniques are triangulation, which approximates each curvilinear feature with several short line segments, and Partial Differential Equation, which is a mesh-free image representation.
We discuss the key benefits and drawbacks of Vector Image representation in Table~\ref{tab:2}.

\begin{table*}[!htbp]
\centering
\caption{Spatial Relation Images Advantages and Disadvantages}
\label{tab:0}
\begin{tabular}{p{0.45\textwidth}p{0.45\textwidth}}
\hline
    \textbf{Advantages} & \textbf{Disadvantages} \\ 
\hline
        \textbf{Storage:} duplicate polygon boundaries are not repeated & \textbf{Not invariant:} properties of geometric features are not invariant under a strict topological definition \\
        \textbf{Error check:} errors introduced during map digitizing and data entry can be automatically checked; in practice, the sum of the space filling polygons of an area is the same as the size of the area that they cover & 
        \textbf{Additional Computational Time and Storage:} topological data structures must be created whether they are used or not; adjacent polygons that share common boundaries
\\ \hline
\end{tabular}
\end{table*}

\begin{table*}[!htbp]
\centering
\caption{Vector Image Advantages and Disadvantages}
\label{tab:2}
\begin{tabular}{p{0.45\textwidth}p{0.45\textwidth}}
\hline
\textbf{Advantages} & \textbf{Disadvantages} \\ 
\hline
    
        \textbf{Easy Manipulation:} easily scaled and otherwise manipulated to accommodate the resolution of a spectrum of output devices.
        & 
        \textbf{Restrictive:} cannot store extremely complex images, e.g., if color info is paramount it varies on a pixel-by-pixel basis
        \\
        \textbf{Easy Conversion:} easy to render data and save it to another format file or vector format, with good results.
        & 
        \textbf{Format Incompatibility:} appearance can vary considerably depending upon the application interpreting the image
        \\
        &
        \textbf{Device Based Quality:} high-resolution raster displays are needed to display vector graphics as effectively.
        \\
        &
        \textbf{Low Performance:} image reconstruction may take considerably longer compared with a bitmap file of equivalent complexity because each image element must be drawn individually and in sequence.
\\ \hline
		\end{tabular}
\end{table*}

Another common structure used in image representation is the Quadtree.
A Quadtree represents an image by recursively splitting it into four distinct quadrants or squares. 
Each node in a Quadtree has a key or a location code/quad code, while the root node symbolizes the first quadrant, which contains the entire image.

A Quadtree representation is efficient when applying some of the basic image processing operations, i.e., union, intersection, comparison, and difference between images, and also it reduces the required storage space~\citep{Kirichek2018QT}.
Several approaches for Quadtree representation have been proposed for specific applications.
Each approach should be used according to its application needs, as none efficiently supports all operations or presents the best compromise solution between compaction and image manipulation.
In Table~\ref{tab:3}, we present the advantages and disadvantages of using a Quadtree image representation, while Table~\ref{tab:4} lists some of the most important Quadtree representations used in applications.

In the medical field, Quadtree is used to efficiently analyze and manipulate images from different medical devices.
\citet{Brindha2023} present an approach to consistently segment images of distinct cell types growing in dense cultures that were captured using various morphological techniques. 
A Quadtree image representation is implemented using the proposed approach that is used to calculate the cell number and density estimates of the tumor from Voronoi tessellation.
\citet{Jewsbury2021} also use a Quadtree representation for computational pathology.
The proposed method generates an interpretable image representation of computational pathology images using Quadtrees and a pipeline to use these representations for highly accurate downstream classification. 
\citet{Hou2020} discuss the challenges of detecting tuberculosis (TB) and proposes a method to enhance TB detection using a different combination of machine learning and image processing methods on the image dataset. 
The proposed solution uses Quadtree to improve the segmentation process of images and to reduce the number of Pixels in a grid.

\begin{table*}[!htbp]
\centering
\caption{Quadtrees Advantages and Disadvantages}
\label{tab:3}
\begin{tabular}{p{0.45\textwidth}p{0.45\textwidth}}
\hline
    \textbf{Advantages} & \textbf{Disadvantages} \\ 
\hline
        \textbf{Easy Conversion:} speeds up the execution time of the applications, provided the quadtree is well chosen  
        &
        \textbf{Restrictive:} depending on the application and on the general operations that need to be performed on the quadtree, it can be very difficult to decide which type to use
\\ \hline
\end{tabular}
\end{table*}

\begin{table*}[!htbp]
\centering
\caption{Quadtrees Applicability}
\label{tab:4}
\begin{tabular}{p{0.15\textwidth}p{0.35\textwidth}p{0.37\textwidth}}
\hline
\textbf{quadtree} & \textbf{Area of application} & \textbf{Limitations} \\
\hline
\begin{tabular}[c]{@{}l@{}}Linear\\ structures\end{tabular} & Intense network transfer & Lack of generality: manages only binary images \\
 & Manipulation of large binary images & Requires additional operations: reading an image consists in undoing it \\
 &  & Requires more resources: a cluster of images is stored by coding each image independently \\
Overlapping structures & Manipulating a large sequence of images & Higher complexity and additional resources: modifications to the sequence requires the creation of new quadtrees that are added at the end of the sequence \\
Inverted structures & Fuzzy search and image querying, large image manipulation,  comparison of homogenous regions for different images & Higher time complexity: Reading an image is more time consuming than using independent quadtrees

\\ \hline
\end{tabular}
\end{table*}

\section{The Proposed Solution}\label{sec:solution}

\begin{figure*}[!ht]
\centering
	\includegraphics[width=.8\textwidth]{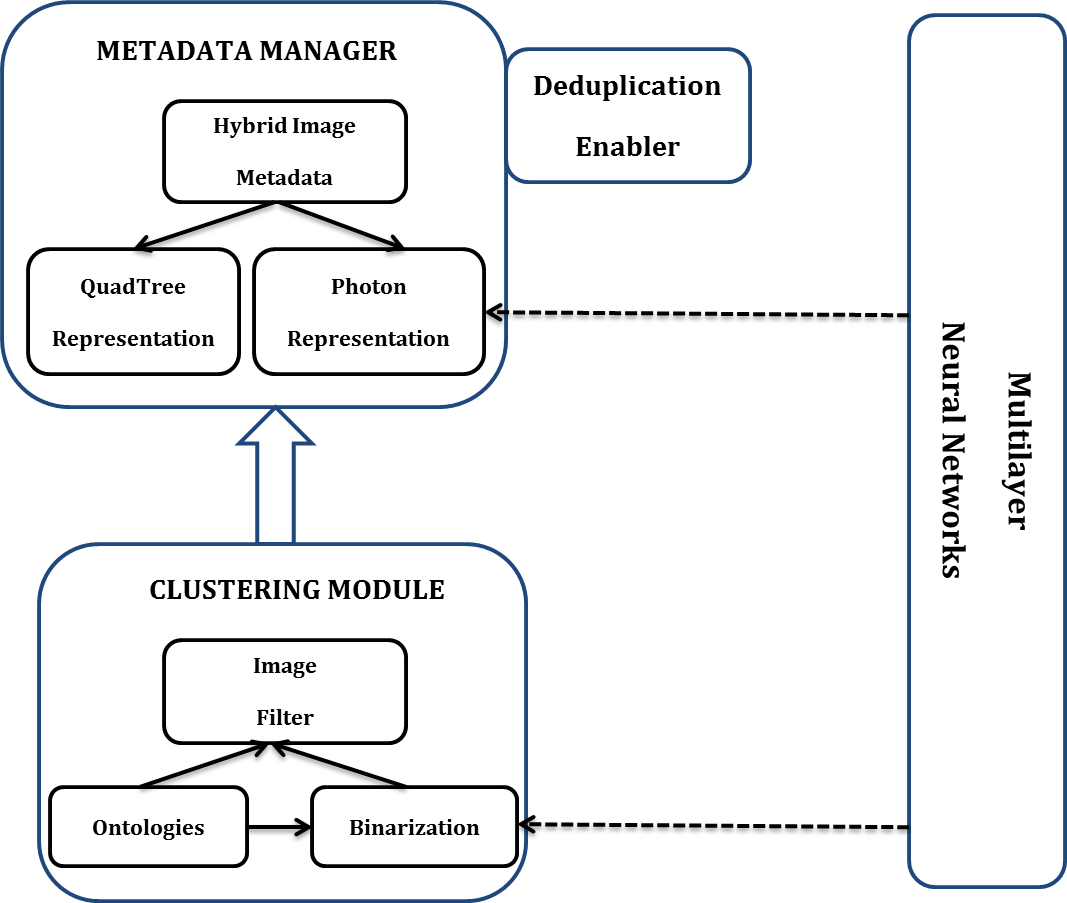}
	\caption{System Representation}
	\label{fig:system_repr}
\end{figure*}

In this section, we present the proposed novel image representation solution and its modules.
The System Representation diagram (Figure~\ref{fig:system_repr}) exposes the interactions between different components of the proposed solution.

Diagnostic imaging services assist medical professionals in interpreting, monitoring, and treating their patients' health conditions.
Referring medical professionals work with diagnostic imaging providers to select the most appropriate and the least invasive diagnostic imaging examination that can achieve the desired results.
Diagnostic examinations include lots of examination procedures such as radiology, magnetic resonance imaging, computed tomography, etc.
In order for diagnostic imaging professionals to interpret the results obtained from different types of images, a standard must be defined.
We want to propose such a standard, by providing a new hybrid image representation for efficient manipulation and processing.
The new representation incorporates metadata information obtained by applying several Machine Learning algorithms (i.e., clustering and neural networks).

\subsection{Clustering Module}
The main goal of the Clustering Module is analyzing and processing the image in order to define its context (region of interest, background), to label the image or different objects in the image, and to improve the image's quality, if possible.
In order to do so, we perform different image processing techniques on a single image or on a group of images.
First of all, we need to identify and locate sharp discontinuities in the image.
The discontinuities are relevant when trying to locate different objects in an image and to establish their positions.
We define them as abrupt intensity changes in pixels that characterize the boundaries of objects.

In order to define the boundaries of different objects, we propose an Image binarization algorithm.
Binary images are often used in digital image processing as masks or as the result of certain operations such as edge detection, segmentation, thresholding, etc.
The global methods use one calculated threshold value to divide image pixels into object or background classes, whereas the local schemes can use many different adapted values selected according to the local area information. 
Hybrid methods use both global and local information to decide the pixel label.
Up to now, the most popular method for image binarization uses the image histogram.
However, selecting the corresponding threshold for each image from different types of applications is still an open issue.
 
After binarization, we are properly determining the edges of each object in the context defined by the image.
The Edge detection algorithm that we explored involves convolving the image with an operator (a filter) that is sensitive to gradients in the image.
Some of the most important variables in defining an edge detection operator are edge orientation, noise environment, and edge structure.
After analyzing different edge detection filters, we identified some of the common problems, i.e., false edge detection, missing true edges, producing thin or thick lines, problems due to noise in the image, etc.
The most commonly used edge detection techniques are Sobel operator, Robert's cross operator, Prewitt's operator, and Laplacian of Gaussian~\citep{Shrivakshan2012}.
In our model, we use the Gaussian operator and a Canny edge detection algorithm, as
the Canny's edge detection algorithm performs better than all the other operators under general scenarios.

In order to better analyze the similarities and differences between a set of images, 
we also propose a new ontology that helps us classify the objects found in an image.
Domain ontologies can be used to provide support in image representation; they can supply added-value information about the structural and logical properties of the model.
When acquiring a medical image, only some of the image characteristics are relevant for a medicinal doctor.
In such cases, ontologies can be used to help the medicinal doctor better frame and classify the object of interest.
The ontologies can be resumed only to the relative location of different organs or can add a qualitative measure such as size, shape, color, etc. Different inferences can be used in order to determine some correlation between the objects of interest and even to establish some environmental influences.

Besides the domain ontology and image representations, we also use K-Means to cluster the images and Fuzzy C-Means to cluster the pixels in order to improve the image's resolution.

Given the binarization representation of a set of $n$ images as $d$-dimensional vectors $(x_1, x_2, \dots, x_n)$, the K-means algorithm objective is to partition the $n$ vectors into $k$ sets $S=\{S_1, S_2, \dots, S_k \}$, with $k \ll n$, that better group the vectors based on some hidden similarity features.
The algorithm works as follows:
\begin{itemize}
    \item[(1)] Determine the initial $k$ centroids $C^{(0)}=\{c^{(0)}_{1}, c^{(0)}_{2}, \dots, c^{(0)}_{k} \}$ either randomly selected points from the dataset or using a heuristic, e.g., K-Means++;
    \item[(2)] Assign each vector $x_{i}$, $i = 1, \dots, n$ to the group defined by the closest centroid: $S^{(t)}_{j} = \{x_{i} : ||x_{i} - c^{(t)}_{j}|| \leq ||x_{i} - c^{(t)}_{j^{*}}|| j^{*} \neq j\}$, where $t$ is the current iteration; 
    \item[(3)] When all vectors have been assigned, recalculate the positions of the $k$ centroids: $c^{(t+1)}_{j} = \frac{1}{|S^{(t)}_{j}|}\sum_{x_{i} \in S^{(t)}_{j}}x_{i}$;
    \item[(4)] Repeat steps (2) and (3) until the centroids no longer move.
\end{itemize}

The Fuzzy C-Means~\citep{suganya2012fuzzy} algorithm work in a similar manner as K-Means.
The main difference between the two algorithms is that Fuzzy C-Means determines the membership level of each point to a cluster instead of assigning the point to a single cluster as in the case of K-Means.
Thus, Fuzzy C-Means returns a set of clusters $C = \{c_{1}, ..., c_{c} \}$ and a partition matrix $W=[w_{ij}]$ ($i = 1, \dots, n$ and $j = 1, \dots, c$), where $w_{ij} \in [0,1]$ tells the membership level of $x_i$ to belong to $c_j$.
To compute $w_{ij}$, a hyperparameter $m$ is introduced to controls the fuzziness of the clusters.
For a set of $n$ images as $d$-dimensional vectors $(x_1, x_2, \dots, x_n)$, the algorithm works as follows:
\begin{itemize}
    \item[(1)] Initialize the partition matrix $W^{(0)}$ at random or using a heuristic ;
    \item[(2)] Compute the centroids for current iteration $t$: $c^{(t)}_{j} = \frac{\sum_{i=1}^{n}w^{m}_{ij}x_{i}}{\sum_{i=1}^{n}w^{m}_{ij}}$
    \item[(3)] Compute the updated weight for the partition matrix $W^{(t+1)}$ for next iteration $t+1$ as $w_{ij} = \frac{1}{\sum_{k=1}^{c}\left( \frac{|| x_{i} - c_{j}||}{|| x_{i} - c_{k}||} \right)^{\frac{2}{m-1}}}$
    \item[(4)] Repeat steps (2) and (3) until $|W^{(k+1)} - W^{(U)}| \leq \varepsilon$, where $\varepsilon$ is a given threshold.
\end{itemize}

\subsection{Metadata Manager}

The Metadata Manager transforms the preprocessed image, obtained from the clusterization module, in the proposed format, and it transmits it across the cloud for further processing.
The hybrid metadata image is obtained by adding the information from a photon map to the information gained after clusterization.
We consider the image as a matrix with different light areas.
Each area has a different light intensity.
A photon represents the smallest particle from the ray of light of the chosen area.
In the first stage of processing, the photon map is built by emitting photons from the light sources.
The photons are stored in the photon maps when they hit non-specular objects.
The second stage, the rendering pass, is used for information extraction about the incoming flux and the radiance reflected by every point in the scene.

The photon map is generated based on the geometric representation of the scene.
The main advantage of photon maps is that no meshes are required.
The error in a photon map is of low frequency, compared to high frequency noise introduced by other algorithms.
The price paid in order to maintain a photon map is the extra memory used to store the photons, especially if the photons are stored in every point of the scene, even where their importance is irrelevant.
A solution is to make the deposition of photons based on their importance, which leads to a more precise photon map and it diminishes the memory required.
Thus the photons are stored only in areas of high visual impact, using an \textit{importon map}.
The importons store color information that describes the percentage with which an illumination at a certain location would contribute to the final image.
When activated, they provide complementary information that can be helpful in the process of deciding how to improve the image's quality given the resources.
They can speed up the lookup of the global illumination photon map.
Another important aspect is the intensity of the importon once it hits an opaque geometry.
In this regard, we defined a model where the importons are blocked and another one, where the importons get stored for all intersections and travel across, losing their intensity power.
Using an importon map increases query speed and reduces memory costs.

Another problem to consider is the automated generation of caustics~\citep{Jonsson2012}.
The photon map used for the caustics requires a much higher resolution.
By extending the approach of driven photon deposition, we will enable the generation of all important caustic effects.
To solve this problem, the photons are divided into two disjoint sets: the set which contains all the photons scattered by a non-diffuse surface, and the rest of the photons map.
Using an approximation function, not only the direct caustics will be covered but also the indirect ones.
The photons, for our representation, are stored using Quadtree which will give us pyramidal access to the image, solving the zooming and rotation problems.

\subsection{Multilayer Neural Networks}

We use a Machine Learning algorithm, i.e., Multilayer Perceptron (MLP) Neural Networks, for (1) selecting the disjoint sets of photons, (2) detecting the name of the organ/tissue, and (3) the binarization process.
An MLP Neural Network is a Deep Learning architecture that stacks multiple layers of fully-connected Perceptron units.
Because the connections between the layers are directed from the input to the output, the MLP model is a feed forward architecture.
For a dataset $X \in \mathbb{R}^{n \times m}$ of size $n$ with $m$-dimensional vectors $x_i \in X$ labeled with a corresponding class $y_i \in Y$, the Perceptron (Equation~\ref{eq:perceptron}) is a simple non-linear processing unit that tries to predict $x_{i}$'s label $\hat{y}_i \in \hat{Y}$ by adjusting a weight vector $\bm{w} \in \mathbb{R}^{m}$ using an activation function $\delta$.
For our MLP model, we use the sigmoid activation $\delta(x)=\frac{1}{1+e^{-x}}$.
The objective for a good prediction is to minimize the average cross-entropy loss function between the prediction $\hat{Y}$ and the true class $Y$ (Equation~\ref{eq:loss_fn}).

\begin{equation}\label{eq:perceptron}
    \hat{y}_i=\delta(\bm{w} \cdot x_i + b)
\end{equation}

\begin{equation}\label{eq:loss_fn}
    L(\hat{Y}, Y) = -\frac{1}{n}\sum_{i=1}^{n}(y_i\log\hat{y}_i + (1 - y_i) \log(1 - \hat{y}_i))
\end{equation}

\begin{figure*}[!htbp]
	\centering
	\includegraphics[width=0.8\textwidth]{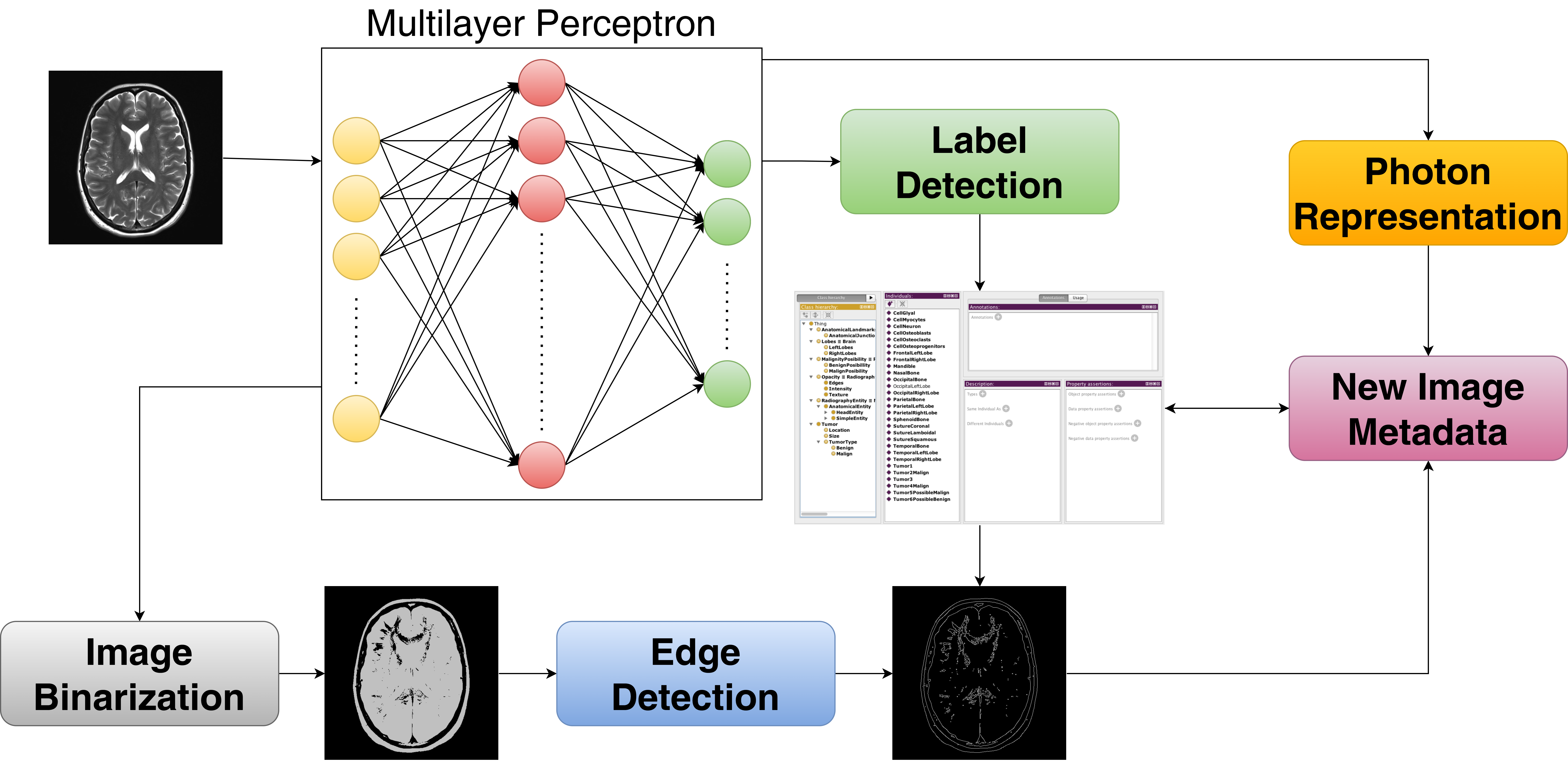}
	\caption{Multilayer Perceptron: Workflow for the new Image Metadata}
	\label{fig:MLP_Diagram}
\end{figure*}

The input of the Multilayer Perceptron is represented by the pixels/photons/importons that need to be classified and the output, the location, or the set where the pixels/photons/importons belong to.
Also, we use of the Multilayer Perceptron in the image classification process, along with the ontology, i.e., the input is represented by an image of an organ/tissue, and the correct output is the name of the organ/tissue.
A single-layer neural network, however, must learn a function that outputs a label solely using the intensity of the pixels in the image.
There is no possibility for it to learn any abstract features of the input since it is limited to having only one layer.
A multi-layered network overcomes this limitation as it can create internal representations and learn different features in each layer.
The first layer is responsible for learning the orientations of lines using the inputs from the individual pixels in the image.
The second layer combines the features learned in the first layer and learns to identify simple shapes such as circles.
Each higher layer learns more and more abstract features such as those mentioned above that can be used to classify the image or the pixels.
Each layer finds patterns in the layer below it, and it has the ability to create internal representations.
The goal and motivation for developing the backpropagation algorithm are to find a way to train the multi-layered neural network such that it can learn the appropriate internal representations of the image and to be able to learn any arbitrary mapping of input to output.
Figure~\ref{fig:MLP_Diagram} presents the workflow used for processing a medical image to obtain the new image metadata.

The purpose of Image Binarization is to extract the lightness (brightness, density) as a feature amount from the image.
Usually, the binarization process is carried out with a threshold found from the histogram of an image automatically.
The threshold deciding criteria is the most difficult part since there are several interfering noises that need to be eliminated.
However, by using the backpropagation network, we can improve the binarization process.
The photons need to be separated into two sets, deciding based on the threshold process, which photons emit light and which photons do not.
Therefore, we can use the multilayer network to achieve that.
The multilayer network is also relevant when trying to define the quantity of energy that the importons will carry once they hit an opaque surface.

\subsection{Deduplication Enabler}

As we use the Cloud environment for storing and executing data, one of the main drawbacks is that different parts of the image need to be transferred to different nodes.
The transfer will increase the computational time (considering the limitations: bandwidth, throughput, etc.) and the price.
Thus, we propose a new transfer method based on data deduplication.
Data deduplication is defined as a data compression technique used for eliminating data copies.
An image consists of many irrelevant data points that will only slow down the image processing algorithm.
We address this problem by defining a Merkle tree structure on top of the Quadtree layers.
For each image, we considered the corresponding chunks from the Quadtree representation.
For all the images, we only maintained the unique chunks, hence we reduced the amount of data stored in the cloud

\begin{figure}[!ht]
	\centering
	\includegraphics[width=0.6\textwidth]{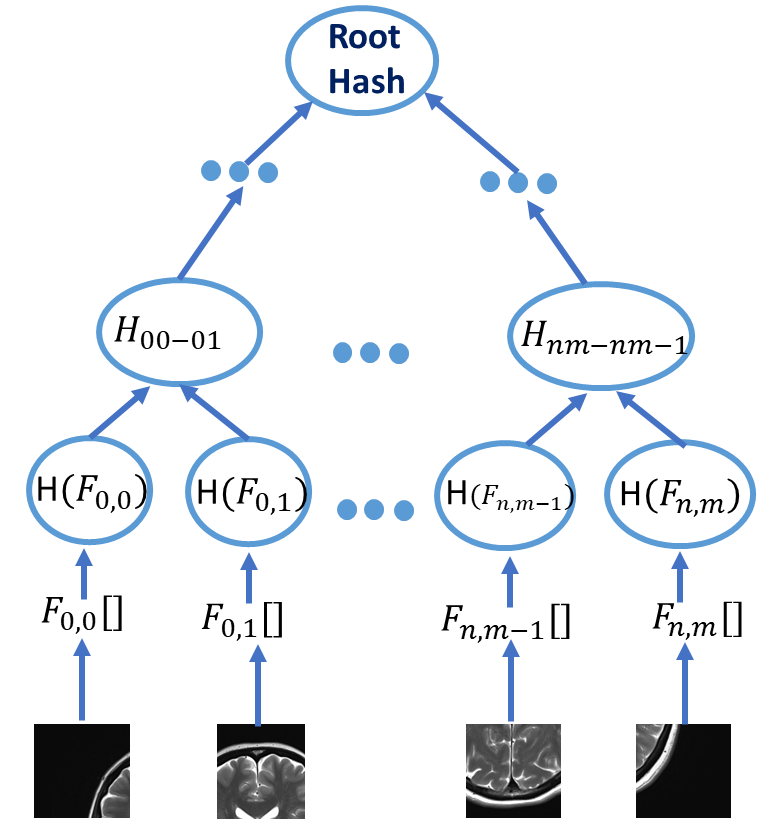}
	\caption{Merkle Tree reprezentation}
	\label{fig:merkle}
\end{figure}

In a Merkle tree, every non-leaf node is labeled with the hash of the labels of its children nodes, while the leaves contain the relevant data.
Taking into consideration that the images and the sections in an image are classified, the Merkle structure allows us to improve the computational time for the transfer.
The classification is important because, for example, for different chunks of the opaque background of an image, we obtain the same hash value.
Therefore, we don't have to transfer those chunks.
If the images or the chunks would not be classified, we would lose time by comparing random data.
Figure~\ref{fig:merkle} presents the structure of the proposed Merkle tree. 
Before constructing the Merkle tree for an image, the point features are extracted.
These feature vectors are used to construct the Merkle tree instead of the images' chunks.
The leaves contain the unique features of each chunk, i.e., $F_{i,j}$.
The value of each non-leaf node is obtained by applying a hashing function to its two children nodes.

Algorithm~\ref{alg:deduplication} presents the deduplication steps with the construction of the Merkle tree. 
The algorithm receives as input the set of medical images $I$ represented using the Quadtrees and outputs the set of Merkle trees $M$ for the images in $I$.
For each image, we get the number of chunks resulting after splitting the image using the Quadtree format (Line~\ref{alg:deduplication:line3}).
Each chunk is identified using a line and column number, e.g., $(i,j)$.
For each chunk, we compute the feature vector (Line~\ref{alg:deduplication:line9}) and the corresponding hash (Line~\ref{alg:deduplication:line11}).

The required steps in case of a transfer request are presented in Algorithm~\ref{alg:transfer}.
The transfer can be done on the same data site or to a remote data site.
If an image must be transferred, send first the image's Root hash (Line~\ref{alg:transfer:line4}).
For each chunk, send the hash value and the feature vector (Lines~\ref{alg:transfer:line5}-~\ref{alg:transfer:line9}).
If a chunk is found at the destination (a chunk with the same feature vector and hash value) then there is no need to transfer that chunk (Lines~\ref{alg:transfer:line10}-~\ref{alg:transfer:line11}). 
Otherwise, send the whole chunk (Lines~\ref{alg:transfer:line13}-~\ref{alg:transfer:line14}).
After a transfer is done, we verify if the image was correctly transferred by comparing the resulting Root hash determine at the destination with the sent one (Algorithm~\ref{alg:reconstruction}). 

\begin{algorithm}[!ht]
\SetKwInOut{Input}{Input}
\SetKwInOut{Output}{Output}
\Input{The set of images $I$}
\Output{The set of Merkle Trees $M$ for the images in $I$ with chunk sizes}

\BlankLine
\tcc{Initialize the set of Merkle Tress $M$}
\emph{$M=\emptyset$} \label{alg:deduplication:line1}

\BlankLine
\tcc{For each image $\varphi$ in the set of images $I$}
\ForEach{$\varphi \in I$}{ \label{alg:deduplication:line2}
    
    \BlankLine
    \tcc{Get the number of chunks from image $\varphi$ Quadtree}
    \emph{$n, m = GetChunkSize(\varphi)$}\; \label{alg:deduplication:line3}
 
    \BlankLine
    \tcc{Get the features set $F$ and the hash set $H$ for the image $\varphi$}
    \emph{$F = \emptyset$}\; \label{alg:deduplication:line4}
    \emph{$H = \emptyset$}\; \label{alg:deduplication:line5}
    
    \BlankLine
    \For{$i = 0, \dots, n $}{ \label{alg:deduplication:line6}
        \For{$j = 0, \dots, m $}{ \label{alg:deduplication:line7} 
        
            \BlankLine
            \tcc{Get the image chunk $\varphi_{i,j}$ }
            \emph{$\varphi_{i,j}  = GetChunk(\varphi, i, j)$}\; \label{alg:deduplication:line8}
            
            \BlankLine
            \tcc{Compute the feature vector $F_{i,j}$ using a feature descriptor}
            \emph{$F_{i,j} = GetFeatures(\varphi_{i,j})$}\; \label{alg:deduplication:line9}
            \emph{$F = F \cup \{ F_{i,j} \}$}\; \label{alg:deduplication:line10}
            
            \BlankLine
            \tcc{Compute the hash for feature $F_{i,j}$}
            \emph{$H_{i,j} = GetHash(F_{i,j})$}\;
            \label{alg:deduplication:line11}
            \emph{$H = H \cup \{ H_{i,j} \}$}\; \label{alg:deduplication:line12}
        }
    }
 
    \BlankLine
    \tcc{Construct the Merkle tree $M_{\varphi}$ for the image $\varphi$}
    \emph{$M_{\varphi} = MerkleTree(F, H)$}\; \label{alg:deduplication:line13}
    \emph{$M = M \cup \{ (M_{\varphi}, n, m) \}$} \label{alg:deduplication:line14}
}    

\BlankLine
\Return{$M$}\;\label{alg:deduplication:line15}
\caption{Deduplication algorithm}\label{alg:deduplication}
\end{algorithm}

\begin{algorithm}[!htbp]
\SetKwInOut{Input}{Input}
\SetKwInOut{Output}{Output}
\Input{A image $\varphi$, \newline
       The set of Merkle Trees $M$, \newline
       A node $\kappa$}
\Output{List of send chunks $\Lambda_{\varphi, \kappa}$}

    \BlankLine
    \tcc{Initialize the list of sent chunks}
    \emph{$\Lambda_{\varphi, \kappa} = \emptyset$}\;\label{alg:transfer:line1}
    
    \BlankLine
    \tcc{Get the Merkle Tree $M_{\varphi}$ for image $\varphi$}
    \emph{$M_{\varphi}, n, m = GetMerkleTree(M, \varphi)$)}\;\label{alg:transfer:line2}
    
    \BlankLine
    \tcc{Get the image root hash $\rho$}
    \emph{$\rho_{\varphi} = GetRoot(M_\varphi)$)}\;\label{alg:transfer:line3}
    
    \BlankLine
    \tcc{Send $\rho_{\varphi}$ to the receiver $\kappa$}
    \emph{$SentTree(\rho_{\varphi}, \kappa)$)}\;\label{alg:transfer:line4}
            
    \BlankLine
    \tcc{Sent the feature vectors $F_{i,j}$ and their hash value $H_{i,j}$ to the receiver $\kappa$}
    \For{$i = 0, \dots, n $}{ \label{alg:transfer:line5}
        \For{$j = 0, \dots, m $}{ \label{alg:transfer:line6}
        
            \BlankLine
            \tcc{Get feature vector $F_{i,j}$  for a chunk from the image's Merkle Tree}
            \emph{$F_{i,j} = GetFeatureVector(M_{\varphi}, i, j)$}\; \label{alg:transfer:line7}
            
            \BlankLine
            \tcc{Get feature vector's the hash $H_{i,j}$ for a chunk from the image's Merkle Tree}
            \emph{$H_{i,j} = GetHash(M_{\varphi}, i, j)$}\; \label{alg:transfer:line8}
            
            \BlankLine
            \tcc{Sent the $F_{i,j}$ and $H_{i,j}$ to the receiver $\kappa$}
            \tcc{The $SentFeatureVectorAndHash$ function return a Boolean acknowledgment if $F_{i,j}$ and $H_{i,j}$ were found or not at the receiver}
            \emph{$exist = SentFeatureVectorAndHash(F_{i,j}, H_{i,j}, \kappa)$)}\;\label{alg:transfer:line9}
            
            \BlankLine
            \If{exist is True}{\label{alg:transfer:line10}
                \tcc{Nothing is send}
                \tcc{Receiver $\kappa$ updates its internal list of chunks $\Lambda_{\varphi}$ with the image chunk $\varphi_{i,j}$ and the list of feature vectors $F'$ 
                }
                \bf{continue}\;\label{alg:transfer:line11}
            }
            \Else{\label{alg:transfer:line12}
            
                \BlankLine
                \tcc{Get the image chunk $\varphi_{i,j}$ }
                \emph{$\varphi_{i,j}  = GetChunk(\varphi, i, j)$}\; \label{alg:transfer:line13}
                
                \BlankLine
                \tcc{Send the image chunk $\varphi_{i,j}$ to the receiver $\kappa$}
                \emph{$SetImageChunk(\varphi_{i, j}, \kappa)$}\; \label{alg:transfer:line14}
                
                \BlankLine
                \tcc{Update the list of send chunks $\Lambda$}
                \emph{$\Lambda_{\varphi, \kappa} = \Lambda \cup \{(i, j)\}$}\; \label{alg:transfer:line15}
                
                \BlankLine
                \tcc{Receiver $\kappa$ updates its internal list of chunks $\Lambda_{\varphi}$ with the sent image chunk $\varphi_{i,j}$}
            }
        }  
    }

\BlankLine
\Return{$\Lambda_{\varphi, \kappa}$}\;\label{alg:transfer:line16}
\caption{Transfer algorithm}\label{alg:transfer}
\end{algorithm}

\begin{algorithm}[!htbp]
\SetKwInOut{Input}{Input}
\SetKwInOut{Output}{Output}
\Input{The root of the Merkle Tree $\rho_{\varphi}$ for image $\varphi$, \newline
       The list of feature vectors $F'$, \newline
       The list of the image chunks $\Lambda_{\varphi}$}
\Output{Reconstructed image $\varphi$}
    \BlankLine
    \tcc{Reconstruct the Merkle Tree $M'_{\varphi}$ from $\Lambda_{\varphi}$ }
    \emph{$F' = \emptyset$}\;\label{alg:reconstruction:line1}
    \emph{$H' = \emptyset$}\;\label{alg:reconstruction:line2}

    \BlankLine
    \ForEach{$F_{i,j} \in F'$}{\label{alg:reconstruction:line4}
        \BlankLine
        \tcc{Recompute the hash for feature vector $F_{i,j}$}
        \emph{$H'_{i,j} = GetHash(F_{i,j})$}\;\label{alg:reconstruction:line5}
        \emph{$H' = H' \cup \{ H'_{i,j} \}$}\;\label{alg:reconstruction:line6}
    } 

    \BlankLine
    \emph{$M'_{\varphi} = MerkleTree(F', H')$}\;\label{alg:reconstruction:line7}

    \BlankLine
    \tcc{Verify if the reconstruction was done correctly}
    \emph{$\rho'_{\varphi} = GetRoot(M'_\varphi)$)}\;\label{alg:reconstruction:line8}
    \If{$\rho_{\varphi} \neq \rho'_{\varphi}$}{\label{alg:reconstruction:line9}
        \Return{\textbf{null}}\;\label{alg:reconstruction:line10}
    }

    \BlankLine
    \tcc{Reconstruct the image}
    \emph{$\varphi = ReconstructImage(\Lambda_{\varphi})$}\;\label{alg:reconstruction:line11}

\BlankLine
\Return{$\varphi$}\;\label{alg:reconstruction:line13}
\caption{Reconstruction algorithm}\label{alg:reconstruction}
\end{algorithm}

The proposed data transfer solution is very efficient when applied to a set of similar images. 
The similarity is established by using the ontology and the clusterization algorithm to classify the images or ROIs (Region Of Interest) from different images before transferring the data to the cloud.

\section{Methodology and Implementation details}\label{sec:implementation}

In this section, we present in detail the proposed algorithms and the implementation steps for our architecture.

\subsection{Clustering Module}

The mechanisms described here are used to represent the medical images.
As a first step, we use the Canny Edge Detection~\citep{Xu2017canny} to filter the noise and to determine the edges.
The algorithm smoothes the image by Gaussian convolution.
The effect of the Canny operator is determined by three parameters, i.e., the width of the Gaussian kernel used in the smoothing phase, and the lower and upper thresholds used by the tracker.
The upper tracking threshold can be set quite high, while the lower threshold can be set low for good results.
We filter the noise from the original image, before the localization and detection of any edges.
We defined a Gaussian filter, and we determine the edge's strength by removing the gradient of the image.
In order to do so, we apply a Sobel operator, that performs a 2D spatial gradient measurement on the image.
We use Equation~\eqref{eq:edge} to find the edge's strength at each point, where $G_x$ and $G_y$ are the gradients obtained by applying two $2 \times 3$ convolution masks.

\begin{equation}\label{eq:edge}
\mid G \mid = \mid G_{x} \mid + \mid G_{y}\mid
\end{equation}

The direction of the edge is computed using the gradient in both directions (Equation~\eqref{eq:edge}).
Once we know the direction, we relate the edge to a direction that can be traced in an image, then we have to resolve the edge orientation in one of the four directions: $0\,^{\circ}$, $45\,^{\circ}$, $90\,^{\circ}$, $135\,^{\circ}$.

\begin{equation}
\Theta = \tan^{-1}\left( \frac{G_y}{G_x} \right)
\end{equation}

We apply non-maximum suppression used to trace along the edge direction and suppress any pixel value.
The final step is used to eliminate streaking.
Streaking represents the breaking of an edge contour caused by the operator fluctuations, above and below a threshold.
By applying the binarization algorithm, we eliminate the streaking.

\begin{figure*}[!ht]
\centering
    \subfloat[Main Entities\label{fig:ontology_entities}]{
	\includegraphics[width=0.49\textwidth]{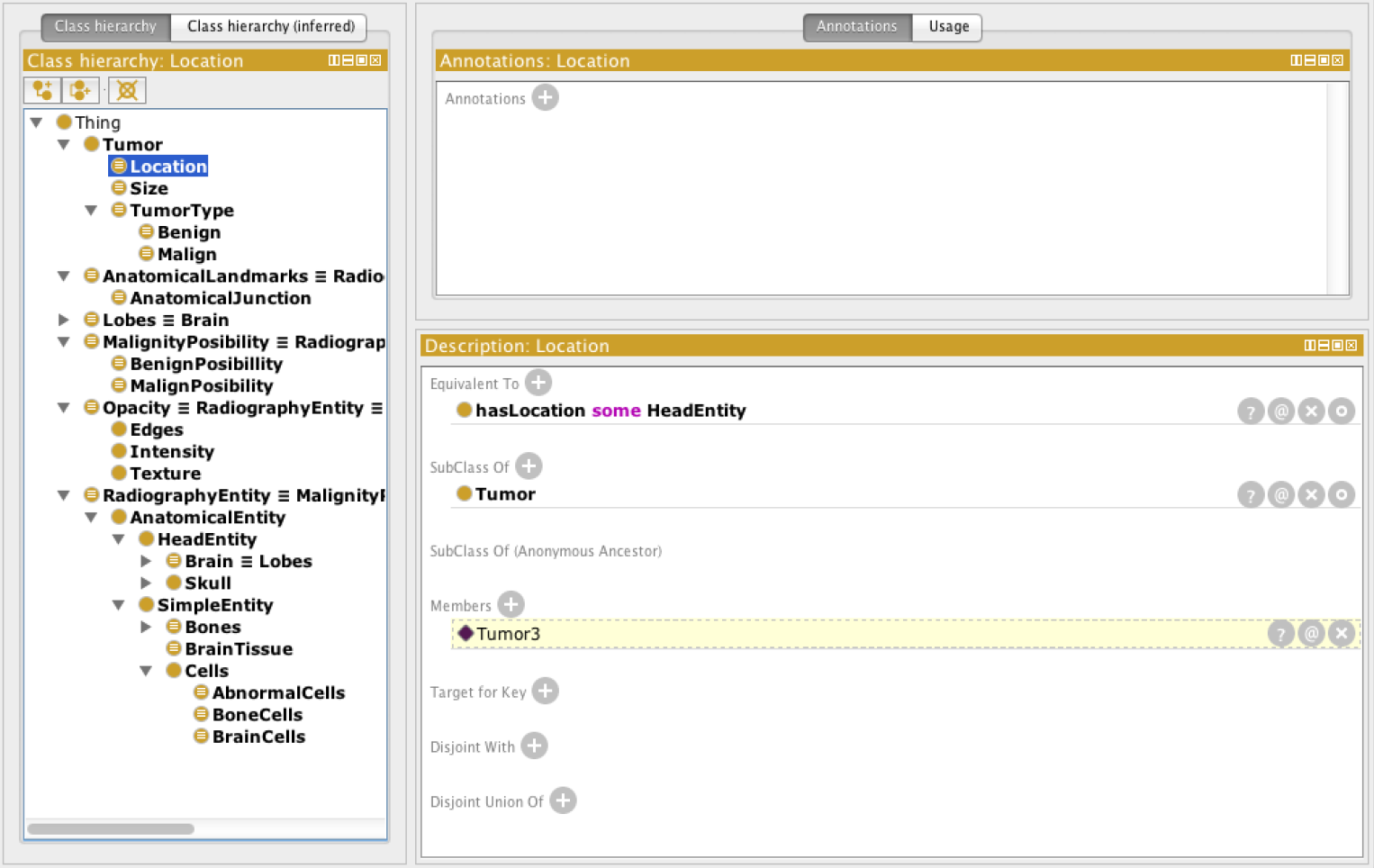}
	}
    \subfloat[Relationships\label{fig:ontology_relationships}]{
	\includegraphics[width=0.49\textwidth]{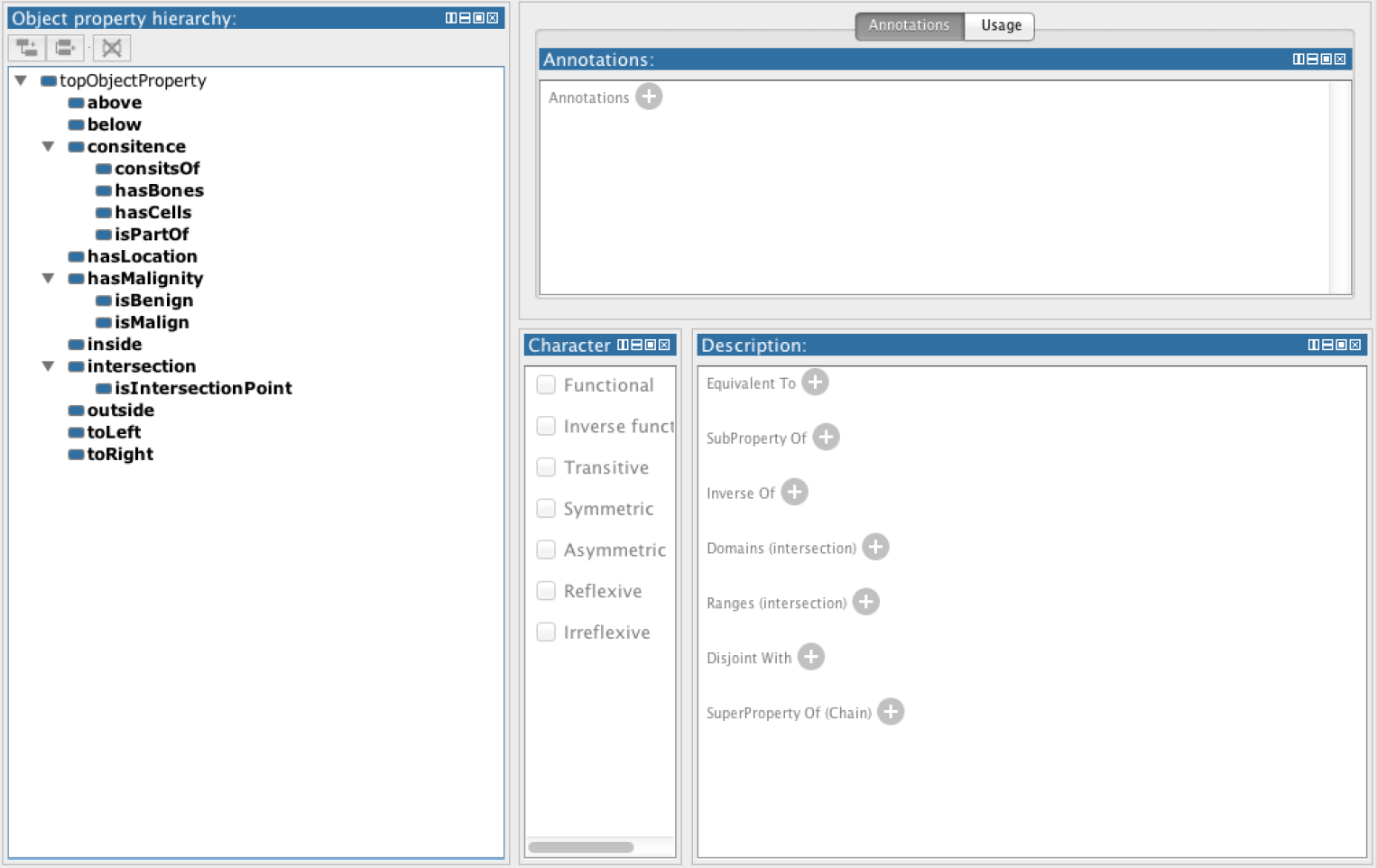}
	}
	\hfill
    \subfloat[Brain Ontology Example\label{fig:ontology_brain}]{
	\includegraphics[width=0.5\textwidth]{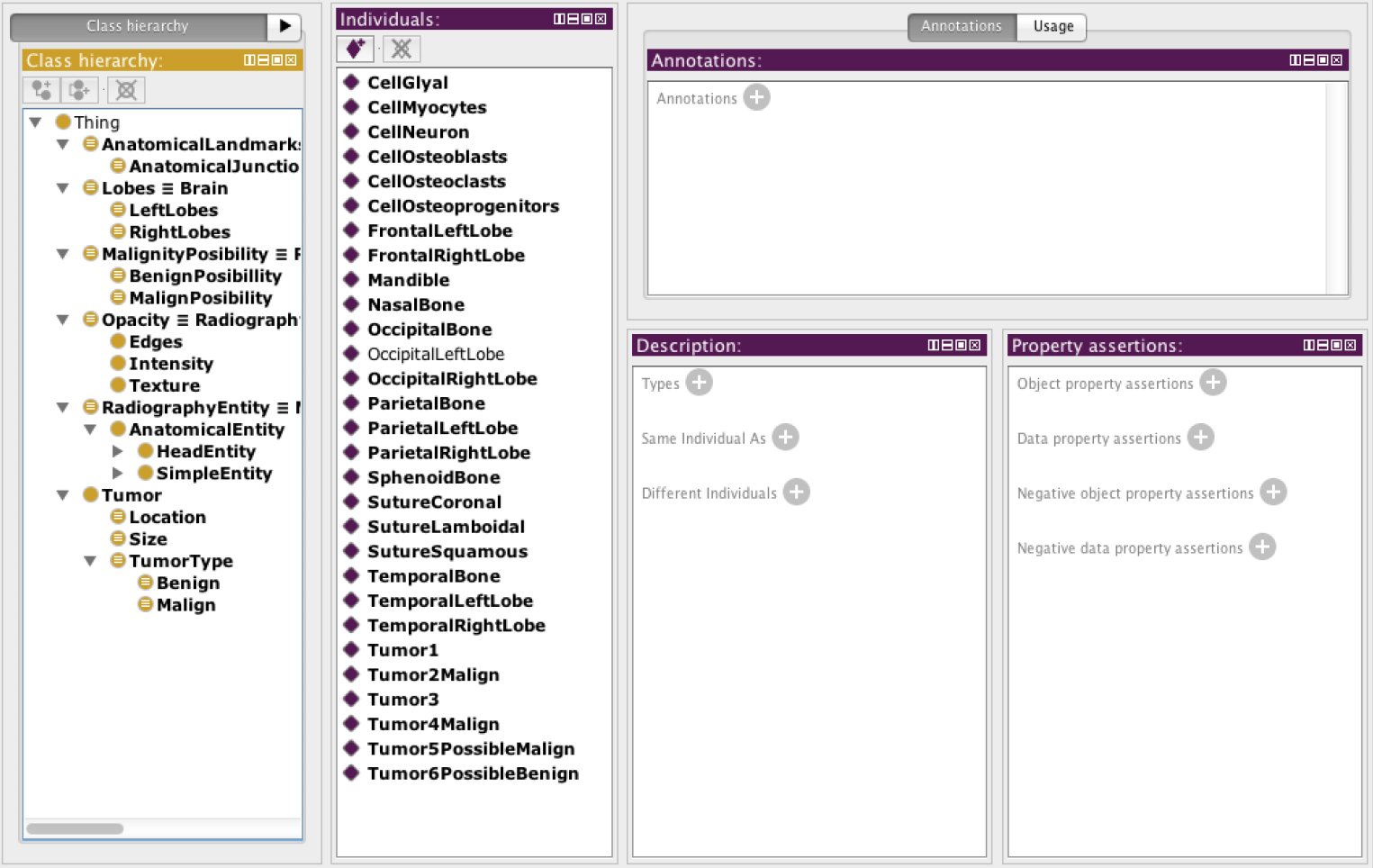}
	}
	\caption{Ontology Representation}
	\label{fig:ontology}
\end{figure*}

In order to improve the images' quality and processing, we introduced an ontology for classifying images or different parts of an image (Figure~\ref{fig:ontology_entities}) and their relationship (Figure~\ref{fig:ontology_relationships}).
When analyzing the image, in order to construct the ontology, we were interested in three major distinct components:
\textit{i)} the particular anatomical structure from which the image is taken;
\textit{ii)} the array of pixels of measured radiation, hydrogen density, etc; and
\textit{iii)} the collection of features each of which is a cluster of pixels with similar pixel values in the pixel array.
Figure~\ref{fig:ontology_brain} presents these components of the human brain.
Before applying the ontology, we analyzed and processed each image.
The first stage of the process of analyzing the image is to cluster pixels into features.
This process is complex, and it must deal with the fact that it is often difficult to extract crisp boundaries from the pixel array, and the pixels that are to be clustered into a single feature may be quite heterogeneous.

\begin{figure*}[!htbp]
	\centering
	\includegraphics[width=1\textwidth]{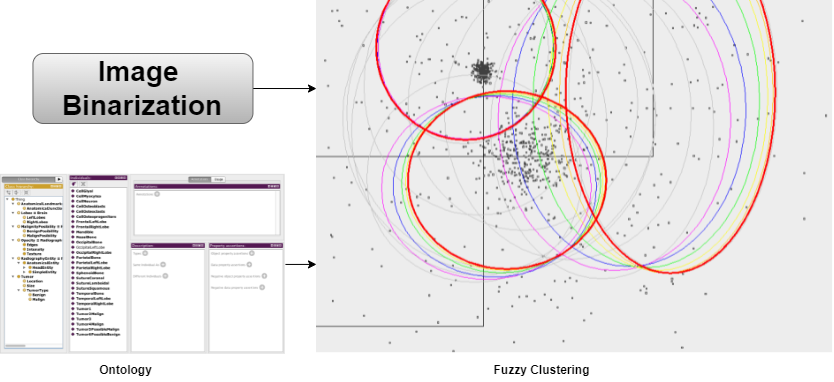}
	\caption{Fuzzy Clustering for feature similarity and boundary detection}
	\label{fig:fuzzy_clustering}
\end{figure*}

In order to cluster the pixels, besides using a fuzzy clustering algorithm and a super-resolution one to determine the boundaries of the image, we used the proposed ontology to analyze the anatomical structures, in a top-down approach (Figure~\ref{fig:fuzzy_clustering}).
The ontology helps us determine which qualitative relations hold between the parts of (normal and pathological) anatomical structures.
After that, we represent the image in the new format, using photons that make it invariant to different transformations.

We also use our ontology to add labels to the different regions in an image.
The labeling system helps us to classify the splitter parts of an image.
After classification, we can apply the proposed deduplication algorithm presented in the previous section.

\subsection{Metadata Manager}

After obtaining the edges of the image, the next step is to convert the image to a grayscale one, if it is in an RGBA (red, green, blue, alpha) representation, but to maintain the original image as well.
Let us consider that the image is represented by a rectangle of $W \times H$, in real space, and each pixel is seen as a $1 \times 1$ rectangle.
In this rectangle, we consider a number of points - photons.
We consider that we have $255$ points in each rectangle that contains pixel $p_{i}$.
We need to establish which pixel is emitting light and which pixel is not.
We will do that by splitting the square: each pixel that will emit or absorb light is strictly related to the neighbors' light intensities.
As a starting point, we can consider that the pixel will emit a different light quantity, that is directly proportional to the neighbor's light emission rate.
Taking into consideration the fact that now we have only the edges of the region of interest (ROI), we can assume that the edges will emit light and that each pixel is a ponder mean of its neighbors: if the pixel is surrounded by emitting pixels, it will be black otherwise, it will have a degree of white in it.
We use the multilayer neural network to define which pixel will emit light and which will not and to modify the quantity of light the pixels emit.
We define the weight of a perceptron in the network as the quantity of light it can emit.
From one pixel to another, the quantity of light emitted diminishes, therefore after traversing the edge, the last pixel should emit a quantity of light close to a chosen epsilon.
In the second phase of the algorithm, we need to update the weight-synapse of the pixels in the image.
The epsilon will influence the speed and the quality of learning which are the pixels that emit light.
We will repeat the algorithm until the performance of the network is satisfactory: all terminal pixels from an edge will emit a quantity of light below epsilon.

We apply MapReduce in order to improve the computational calculus.
The mappers read parts of the preprocessed image, i.e., the image represented in gray scale, with a canny edge.
Each mapper sends to the reducer a line representing the pixel values from the image.
The reducer calculates the ponder mean and applies the neural network algorithm and sends back to the mapper the value to be written in the original image.
The mapper changes the values for the image.

We apply the metadata gained from the ontology, canny edge, and photon representation for each pixel in the original image.
After that we split the image based on the classification obtained from the ontology: we can use different objects identified in the image or other criteria.
The final image is represented in a Quadtree.

\subsection{Deduplication Enabler}
The Deduplication Enabler improves the transfer rate in the distributed environment.
When defining the deduplication method we took into consideration three important aspects: how to chunk the file efficiently, how to better leverage potential similarity and identity among dedicated applications, and how to store chunks effectively and reliably in secondary storage devices.
For our data type, we determine the feature vector for each chunk and use this information and the coordinates of the chunks to construct the Merkle tree.

To detect the points of interest and create the feature vectors, we employ the ORB feature descriptor~\citep{Rublee2011orb}.
Oriented FAST and Rotated BRIEF (ORB) is a feature detector and descriptor used in computer vision. It is used in tasks such as image classification, object recognition, or 3D reconstruction.
ORB combines the functionality of the FAST(Features from Accelerated and Segments Test) keypoint detector and the BRIEF(Binary robust independent elementary feature) descriptor, providing good performance and low cost in comparison with other existing solutions. 

To construct the hash for each node from the tree, we use the Tiger Hashing function and store its values in the Merkle tree.
Tiger~\citep{Anderson1996tiger} is a cryptographic hash function that produces a 192-bit hash value, by processing 512-bit blocks.
It is designed using the nearly universal Merkle-Damgard paradigm, which is a method of building collision-resistant cryptographic hash functions from collision-resistant one-way compression functions. 
The hash function consists of two parts: the key schedule and the state update transformation. 
The first step of the hashing function is to operate on 64-bit words,  maintaining 3 words of state and processing 8 words of data. 
The next main step is composed of 24 rounds, using a combination of operation mixing with XOR and addition/subtraction, rotates, and S-box lookups. 
In the second phase, an intricate key scheduling algorithm is used for deriving 24 round keys from the 8 input words.

The algorithm is applied to a set of similar images.
The similarity is established by using the proposed ontology and a clusterization algorithm. 
Thus, we classify the images or regions of interest from different images before transferring the data to the cloud.

\section{Experimental results}\label{sec:experiments}

In this section, we present the dataset used for evaluating our proposed solution and the experimental results.

\begin{figure*}[!htbp]
    \centering
	\includegraphics[width=0.6\textwidth]{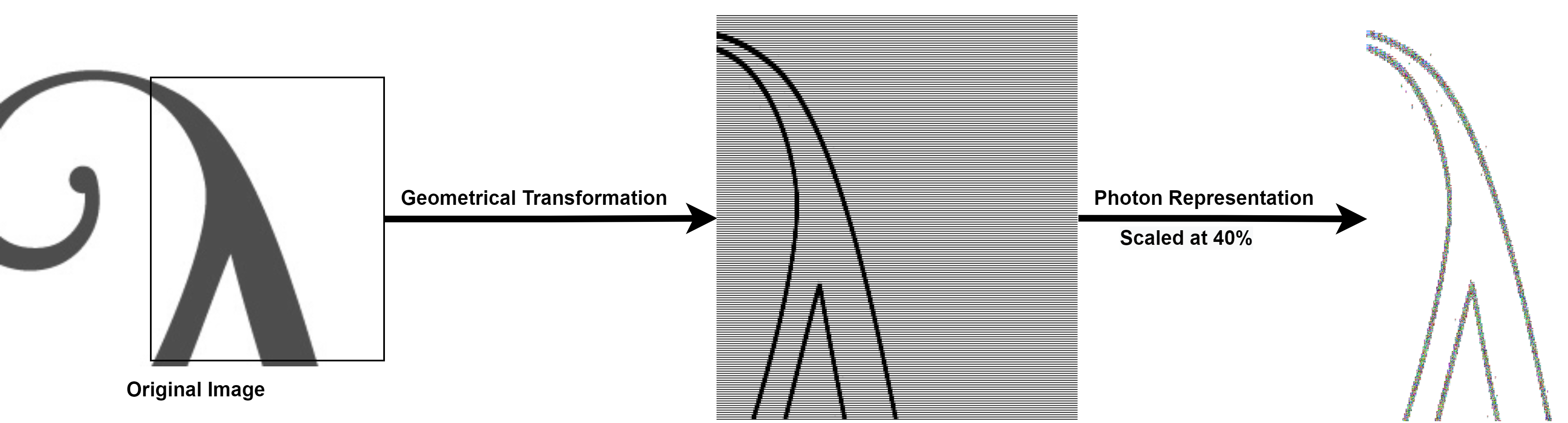}
    \caption{From original image to the photon representation - An example}
    \label{fig:image}
\end{figure*}

\subsection{Dataset and Image Representation}
The dataset consists of real-world anonymized CT (Computer Tomography) images of different cross-sections of the human body.
It has a size of $\sim$ 80 GB.

We apply our algorithms in order to represent the images using our proposed format based on Quadtrees.

Starting from the original image (Figure~\ref{fig:image}), we apply a geometric transformation technique that includes an edge detection algorithm, a binarization algorithm, and a multi-layer network algorithm and obtain a new image representation.
The new image representation has a better resolution than the initial image.
The final step is to represent the image using only the photons.

Next, we used K Means to cluster the images and Fuzzy K Means to cluster the pixels in
order to improve the image’s resolution.
In order to cluster the pixels besides using our photon representation and a fuzzy clustering algorithm, our proposed ontology can be used in order to analyze the anatomical structures as a top-down approach to determine which qualitative relations hold between the parts of (normal and pathological) anatomical structures.

\subsection{Results}

We test our proposed solution in a distributed multi-cloud environment.
For this, we consider two data centers from Microsoft's Windows Azure cloud, i.e., in the North-Central US and North-Europe data centers.

For our experiments, we use a geographical-distributed set-up consisting of small Av2-series Virtual Machines having 1 CPU, 2GB of memory, 500GB local storage, and 100MB/s bandwidth deployed on 8 physical nodes.
It is important to mention that the sender and the receivers are in different geographic data centers.
We set the sender in the North-Europe datacenter and the receivers in the North-Central US.

\begin{figure}[!htbp]
	\centering
	\includegraphics[width=0.75\textwidth]{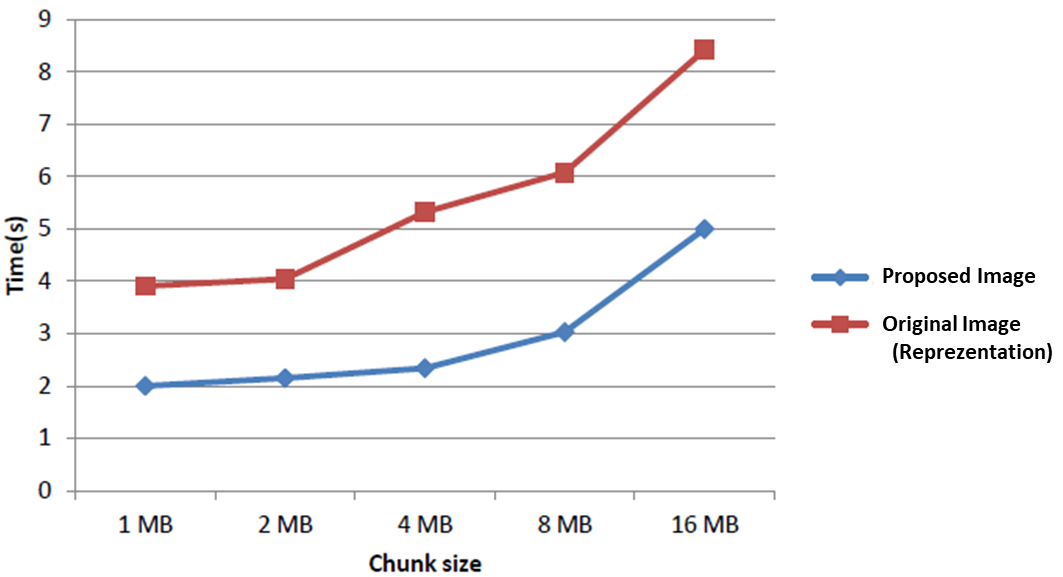}
	\caption{Chunk size variation: Original Image Representation vs Our Solution}
	\label{fig:chunk}
\end{figure}

The medical images represented using our proposed solution are deduplicated and divided into chunks.
The chunks are distributed across different nodes to improve the computational time.
If a node requires an image that is not in its data center, the chunks of that image must be transferred between the data centers.
We first analyze the impact of the chunk’s size on the transfer rate and on the conversion computational time.
Figure~\ref{fig:chunk} presents the chart for the conversion time from the initial image to the photon image for our proposed image in comparison with the original image representation.
Thus, when considering sending data over the Internet, the smaller the chunk, the better the transfer rate.

\begin{figure}[!htbp]
	\centering
	\includegraphics[width=0.75\textwidth]{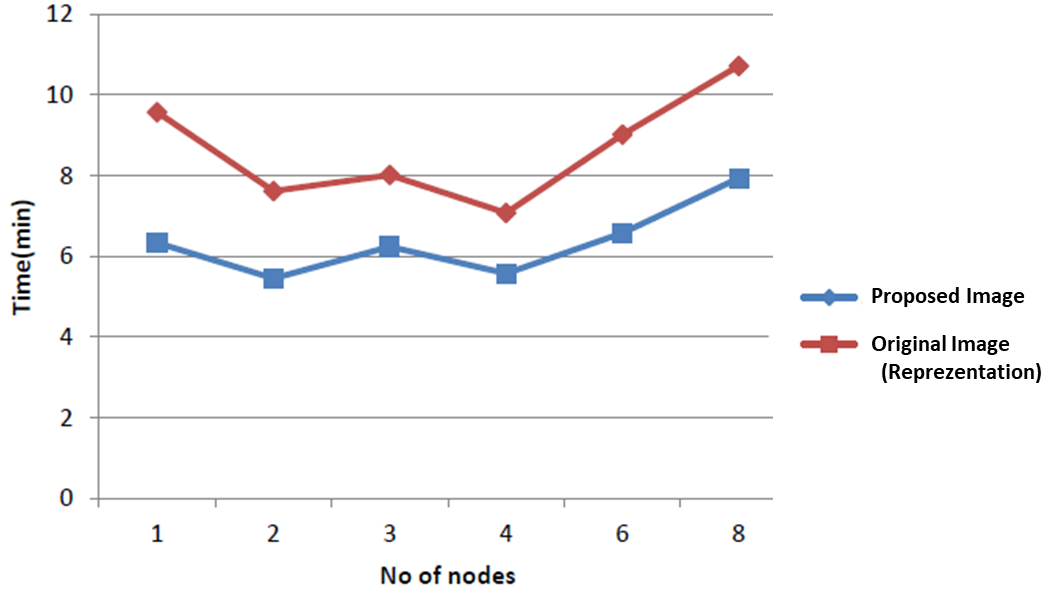}
	\caption{Inter-Cloud Data Transfer: Original Image Representation vs Our Solution}
	\label{fig:merkleimg}
\end{figure}

Our proposed Merkle Tree-based representation obtains on average $27\%$ time improvement when transferring the data (Figure~\ref{fig:merkleimg}).
The data clustering algorithm is performed on the sender machine, before sending the data.

Thus, although at the metadata level, we maintain more information that increases the image size, the data transfer is done in less time.
One important factor is that only unique data chunks are considered in our image representation.
This is very significant, especially in the case of medical images where we can have many image blocks that contain only one color (considered as background).

\section{Conclusions}\label{sec:conclusions}
In this paper, we propose a solution that would unify the representation for the various devices used in acquiring medical images --- answering research question \textit{$Q_1$}.
The new image representation has a better resolution than the initial image and maintains information that has an impact on the diagnosis --- answering research question \textit{$Q_2$}.
Furthermore, using the photons algorithm solves the resolution problem, without the need for post-processing techniques.
As we are dealing with a large amount of data and metadata, we are using a distributed environment and the MapReduce paradigm to speed up the algorithms.
We also proposed and tested a deduplication algorithm that improves the time transfer in the cloud, based on Merkle trees.
Initially, we clustered the images.
After that, we tested the deduplication method by converting the nodes from the Quadtrees to different sets of bytes and applying a hashing function.
In order to obtain a better representation of data in the cloud, we decided to use the feature vectors obtained by applying the ORB feature detector as the first level hashing function for the data blocks and the Tiger Hashing for the following levels.
We maintained a history of the chunks that we already sent, and for each new chunk, we compared its ORB features with the ones we maintained.

In future work, we will analyze different hashing methods and transfer strategies in order to further improve the deduplication method.



\end{document}